\documentclass[a4paper, 10pt]{article}
\usepackage[utf8]{inputenc} 
\usepackage{amsmath}
\usepackage{graphicx}
\usepackage[small,bf]{caption2}  
\usepackage{float}
\usepackage{hyperref}

\begin{document}

\date{July 25, 2023} 
\title{A transfer matrix for the input impedance of weakly tapered, dissipative cones as of wind instruments}
\author{
Timo Grothe$^1$, Johannes Baumgart$^2$, Cornelis J.Nederveen$^3$\\
$^1$ \emph{\small Erich-Thienhaus Institut, Hochschule f\"ur Musik Detmold, 32756 Detmold}
\\
$^2$ \emph{\small Independent Researcher, Dresden, Germany}  \\
$^3$ \emph{\small Independent Researcher, Pijnacker, The Netherlands}  
}
\maketitle

\begin{abstract}
A formula for the local acoustical admittance in a conical waveguide with viscous and thermal losses given by Nederveen [(1969) \emph{Acoustical Aspects of Woodwind Instruments} (Frits Knuf, Amsterdam)] is rewritten as an impedance transmission matrix.
Based on a self-consistent approximation for the cone, it differs from other one-dimensional transmission matrices used in musical acoustics, which implicitly include the loss model of a cylinder.
The resonance frequencies of air columns calculated with this transmission matrix are in better agreement with more comprehensive models. 
Even for long cones with a slight taper, there is no need to discretize along the axis.
\end{abstract}

\maketitle

\section{\label{sec:1} Introduction}
The first mention of a transfer matrix to describe sound propagation in a conical waveguide has been given by Benade~\cite{Benade1988}. 
His derivation from equivalent electrical networks is an analytical solution for the acoustic input impedance of a conical waveguide without losses~\cite{Chaigne2016, Fletcher1988}.
To include the effect of wall losses in the impedance description, it is a common practice to locally apply the thermo-viscous theory of wave propagation in a cylindrical duct. 
This leads to an integral mean value of the propagation constant~\cite{Kulik2007, Walstijn2002} or, equivalently, to an effective radius of the cone accounting for visco-thermal losses at the wall in a single value.
The accuracy required in musical acoustics has recently inspired some studies that reconsider this approach~\cite{Chabassier2019, Ernoult2020, Grothe2013, Thibault2023}.\\
In this Letter, we recall one of the first approaches to treat visco-thermal losses in a cone by Nederveen~\cite{Nederveen1969}.
It is rewritten here in current common nomenclature, brought to the form of a transfer matrix and applied to long, slim cones with typical dimensions of woodwind air columns. 
Results of other transfer matrices and an up-to-date finite-element model are compared. 
\section{Equations}
Nederveen's description [Eq.(23.30) in Nederveen~\cite{Nederveen1969}] 
of the local admittance in a conical waveguide with viscous and thermal losses reads
\begin{equation}
Y(\omega,x)	= \frac{1}{\mathrm{j}\,Z_{c}}\left[\frac{1-\alpha_g}{k\,x}-\frac{1+\alpha_f}{\tan{(k\,x(1+\alpha\,\xi) + \Psi)}}\right]
\label{eq:Y_loss}
\end{equation}
where $\omega$ is the angular frequency, j $= \sqrt{-1}$, $x$ is the axial distance from the cone apex, and $k=\omega/c$ is the wavenumber. $Z_c = \rho\,c/S$ is the local characteristic impedance of the duct with cross section $S(x)$. $\Psi$ is a phase constant determined by the boundary conditions at the ends of the cone. The visco-thermal losses in the fluid at the wall are included in the form of three dimensionless, complex-valued parameters related to the local radius $r(x)$ and the apical distance $x$ as
\begin{subequations}
\begin{eqnarray}
\alpha(\omega,x) &=&C_v \left(\frac{\gamma-1}{\sqrt{Pr}}+1\right)\label{eq:alpha}
\\
\alpha_f(\omega,x) &=&C_v\left(2-2\,f\,k\,x \left(\frac{\gamma-1}{\sqrt{Pr}}+1\right)\right)\label{eq:af_cone}
\\
\alpha_g(\omega,x) &=&C_v\left(1+2\,g\,(k\,x)^2 \left(\frac{\gamma-1}{\sqrt{Pr}}+1\right)\right)\label{eq:ag_cone}
\\
\textrm{with}& & C_v = \frac{1}{2}\sqrt{2}\frac{\delta_v}{r(x)}(1-\mathrm{j})\, \notag
\end{eqnarray}
\end{subequations}
where $\delta_v = \sqrt{\eta/(\omega\rho)}$ is the viscous boundary layer thickness, $\eta$ is the dynamic viscosity, $\rho$ is the density, $\gamma$ is the ratio of specific heats, and $Pr$ is the Prandtl number. 
The functions $f$ and $g$ are evaluations of sine and cosine integrals
\begin{subequations}
\begin{eqnarray}
f(\omega,x)&=&  \mathrm{Ci}(2\,k\,x)\sin(2\,k\,x) - \left(\mathrm{si}(2\,k\,x)-\frac{\pi}{2}\right)\cos(2\,k\,x), \label{eq:f}
\\
g(\omega,x)&=& \mathrm{-Ci}(2\,k\,x)\cos(2\,k\,x) - \left(\mathrm{si}(2\,k\,x)-\frac{\pi}{2}\right)\sin(2\,k\,x). \label{eq:g}
\end{eqnarray}
\label{eq:fg}
\end{subequations}
The scaling coefficient $\xi = \mathrm{ln}(k\,x) + g(\omega,x)$ captures the effects of boundary layers on the propagating wave, namely, apparent mass change and dissipation, as represented by the real and imaginary part of $\alpha$. As $\Psi$ is a constant independent of the spatial coordinate $x$, it cancels out~\cite{Nederveen2008} when writing an admittance transmission equation for a cone of length $L = x_2-x_1$, as
\begin{equation}
Y_1(\omega) = \frac{1}{\mathrm{j}\,Z_{c1}}\,\left[\frac{1-{\alpha}_{g1}}{k\,x_1}- \frac{\left(1+{\alpha}_{f1}\right)\left(\frac{1-{\alpha}_{g2}}{k\,x_2}-\mathrm{j}\,\frac{Z_{c2}}{Z_2}\right)+\left(1+{\alpha}_{f1}\right)
\left(1+{\alpha}_{f2}\right)\tan(\sigma)}
{\left(1+{\alpha}_{f2}\right)-\left(\frac{1-{\alpha}_{g2}}{k\,x_2}-\mathrm{j}\,\frac{Z_{c2}}{Z_2}\right)\tan(\sigma)}\right].
\label{eq:A_Nederveen}
\end{equation}
where
\begin{equation}
\sigma = k\left( x_2(1+(1-\mathrm{j})\alpha_2\,\xi_2)-x_1(1+(1-\mathrm{j})\alpha_1\,\xi_1) \right)
\label{eq:sigma}
\end{equation}
Here, subscripts $(\cdot)_{i}$ with $i = 1,2$ denote input and output end of the cone, respectively.
The input impedance $Z_1$ is the inverse of Eq.~(\ref{eq:A_Nederveen}), which is written as \begin{equation}
Z_1 = \frac{Z_2\,A+B}{Z_2\,C+D},
\label{eq:impedance}
\end{equation}
describes the impedance transformation between output and input end in the form of a transfer matrix $[A,B;C,D]$ with these elements:
\begin{equation}
\begin{array}{lclll} 
A & = & 		           &\hspace{.25em}\frac{r_2}{r_1}     &\cos(\sigma)(1+\alpha_{f2})-\frac{1}{k\,x_1} \sin(\sigma)(1+\alpha_{g2})\\
B & = &Z_{c1}          &\hspace{.25em}\frac{r_1}{r_2}\,\mathrm{j}  &\sin(\sigma)\\
C & = &\frac{1}{Z_{c1}}&[\frac{r_2}{r_1}\,\mathrm{j} &\sin(\sigma)(1+\alpha_{f1})(1+\alpha_{f2})+ ...\\
  &  &                 & &\mathrm{j}\ \frac{1}{k\,x_1}\left(\frac{1}{k\,x_1}\sin(\sigma)(1+\alpha_{g1})(1+\alpha_{g2}) + \right. ...\\ 
  &  &                 & &\hspace{4em} \left. \cos(\sigma)(1+\alpha_{f1})(1+\alpha_{g2})- \right. ...\\
	&  &                 & &\hspace{4em} \left. \frac{r_2}{r_1}\cos(\sigma)(1+\alpha_{f2})(1+\alpha_{g1})\ \right)]\\
D & = &                &\hspace{.25em}\frac{r_1}{r_2} \left(\vphantom{\frac{1}{k\,x_1}} \right.&\left.\cos(\sigma)(1+\alpha_{f1})+\frac{1}{k\,x_1}\sin(\sigma)(1+\alpha_{g1})\right),
\end{array}
\label{eq:TM_Nederveen}
\end{equation}
where $\sigma$ is a complex-valued propagation constant
\begin{equation}
\sigma = k\,L(1+\frac{\alpha^\prime}{r_{\mathrm{eff}}}(1-\mathrm{j})), \label{eq:eqa} 
\end{equation}
which encompasses losses in terms of the ratio of a visco-thermal boundary layer thickness, $\alpha^\prime$, as
\begin{equation}
\alpha^\prime= \frac{1}{2}\sqrt{2}\,\delta_v\left(\frac{\gamma-1}{\sqrt{Pr}}+1\right)
\end{equation}
to an \emph{effective radius} of the conical frustum, $r_{\mathrm{eff}}$, as
\begin{equation}
r_{\mathrm{eff}}      = \frac{r_2-r_1}{g_2-g_1+\ln\frac{r_2}{r_1}}. \label{eq:reff} 
\end{equation}
The main result of this Letter is a transfer matrix for the cone with losses Eq.~(\ref{eq:TM_Nederveen}). It is written here 
following the notation of Benade's transfer matrix for the cone without losses~\cite{Benade1988, Chaigne2016}.
 
\section{Discussion}
The reorganization of Nederveen's admittance formula following the notation of standard transfer matrices can conveniently be compared to the literature.
The contribution of visco-thermal losses is scaled by a dimensionless number which relates the corresponding boundary layer thickness to a characteristic length.
The characteristic length in a duct is the quotient of area and perimeter~\cite{Pierce2019}-- the hydraulic radius. 
The radius change in a cone motivates the definition of an effective radius $r_{\mathrm{eff}}$, which captures the entire losses within the cone in a single value. It is an ansatz to avoid the solution of the underlying equations, recently elaborated by Thibault~\emph{et al.}~\cite{Thibault2023}, that requires numerical iterative methods.\\
Several definitions for an effective radius have been suggested in the past:
The simplest approximation is based on the arithmetic mean $r_{\mathrm{eff}} = (r_2+r_1)/2$.
Another option is to use the radius of a cylinder with the same volume-to-surface ratio as the cone, resulting in $r_{\mathrm{eff}}=(r_2^3-r_1^3)/(3r_1(r_2+r_1))$, which is similar to an empirically useful value~\cite{Chabassier2019} $r_{\mathrm{eff}}= (2\,r_1+r_2)/3$.
A different approach is to calculate an averaged propagation constant for the cone as the integral mean of the local propagation constant from cylinder theory~\cite{Kulik2007, Walstijn2002}, as 
\begin{equation}
\begin{array}{lcl} 
\bar{\sigma}_{con.} &=& \left(\frac{1}{x_2-x_1}\int_{x_1}^{x_2}{k_{cyl.}(x)\,dx}\right) L \\
                    &=&k\,L\left(1+\alpha^\prime(1-\mathrm{j})\frac{1}{x_2-x_1}\int_{x_1}^{x_2}\frac{1}{r(x)} dx\right),\\
\end{array}
\label{eq:sigmacon}
\end{equation}
which, with Eq.~\ref{eq:eqa},  yields $r_{\mathrm{eff}} = (r_2-r_1)/\ln(r_2/r_1)$.
Taking into account that the hydraulic radius in a conical duct refers to a spherical cap area, rather than to a planar cross sectional area, leads to a corrected equivalent radius $r_{\mathrm{eff,sph}} =c_m r_{\mathrm{eff}}$~\cite{Thibault2023}, with the correction factor $c_m = 1+(1/m^2(1-\sqrt{m^2+1})^2)$, depending on the taper $m = (r_2-r_1)/L$ of the cone. 
Nederveen's approach leads to yet another value of $r_{\mathrm{eff}}$, which takes cone length and frequency into account [Eq.~(\ref{eq:reff})].\\ 
Aside from this modified effective radius, two additional dimensionless loss parameters $\alpha_f$ and $\alpha_g$ are introduced, which raise complex-valued scaling factors to the propagation terms in $\sin(\sigma)$ and $\cos(\sigma)$.\\ 
In the limit of negligible losses ($\eta \rightarrow 0$) Eq.~(\ref{eq:TM_Nederveen}) converges to the transfer matrix of a cone without losses ($\alpha = \alpha_f = \alpha_g = 0$), including the limit case of a loss-free cylinder ($r_2\rightarrow r_1$ and $1/(k\, x_i)\rightarrow 0$).
For vanishing taper, however, the limits $\lim_{r_2\to r_1} \alpha_{f,g}$ are small non-zero values in the order of $\alpha$ which prevent convergence towards the transfer matrix of a cylinder with wall losses.
To still achieve convergence in practical applications, it is sufficient to force $\lim_{r_2\to r_1} \alpha_{f} = 0$, e.g., by sigmoid regularization of Eq.~(\ref{eq:af_cone}).\\ 
Nevertheless, it can be shown that the expressions in Eq.~(\ref{eq:TM_Nederveen}) are self-consistent. Kulik~\cite{Kulik2007} has discussed this aspect for a cone transfer matrix extended by the loss model [Eq.~(\ref{eq:sigmacon})], by showing that $A_{0}/A_{n} = B_{0}/B_{n} = C_{0}/C_{n} = D_{0}/D_{n} = 1$ holds regardless of $n$, where the subscripts $(.)_j$ with $j = 0,n$ denote the number of subdivisions of the cone.
Similarly, with $\tilde{Z}_i = Z_i/Z_{ci}$ and $Z_{ci} = \rho\,c/(\pi r_i^2)$ inserted in Eq.~(\ref{eq:impedance}), the matrix elements in Eq.~(\ref{eq:TM_Nederveen}) become non-dimensional and numerical evaluation shows that $A_{0}/A_{n} = B_{0}/B_{n} = C_{0}/C_{n} = D_{0}/D_{n}$. 
This result implies that with Nederveen's cone model, a discretization into shorter conical segments does not affect the impedance result.\\
\section{Application}
\begin{figure}[t]
\includegraphics[width=1\textwidth]{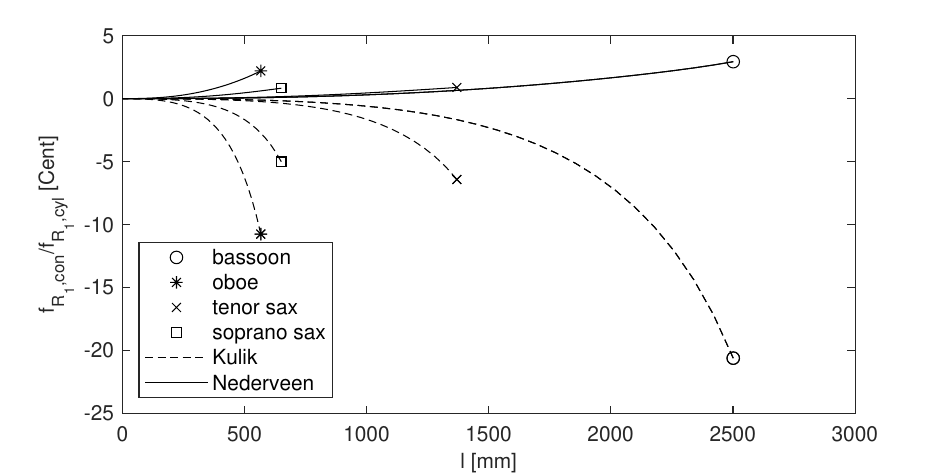}
\caption{\label{fig:FIG1}{Deviation of impedance peak frequency $f_{R1}$ of weakly conical frustums as calculated from different single-cone transfer matrices: Kulik~\cite{Kulik2007} (dashed lines) and Nederveen~\cite{Kulik2007} [(Eq.~(\ref{eq:TM_Nederveen}))] (straight lines), referenced to the result 
of a \emph{cylindrical-slices model}. Markers indicate typical cone geometries of woodwind instruments (see Table~\ref{tab:woodwinds}). Along each curve, the taper $m$ is constant and the input radius $r_1$ varies with $l$ as $r_{1} = R_1+m(L-l)/2$}}
\raggedright
\end{figure}

\begin{figure}[t]
\includegraphics[width=1\textwidth]{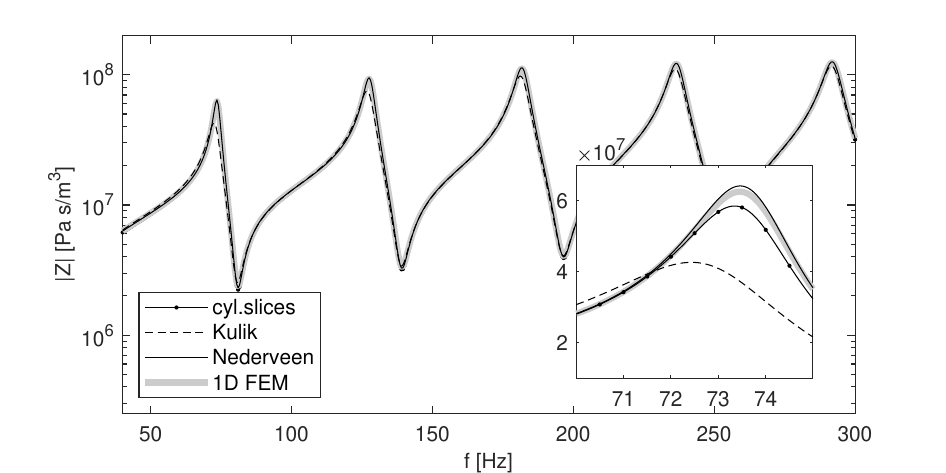}
\caption{\label{fig:FIG2}{Impedance magnitude of a bassoon-like conical frustum ($r_1 = 2.1$~mm, $r_2= 23.5$~mm, $L = 3$~m) closed at the far end. Calculations with three different transfer matrix methods: cylindrical slices, Kulik~\cite{Kulik2007}, and Nederveen~\cite{Nederveen1969} [(Eq.~(\ref{eq:TM_Nederveen}))] in comparison to a numerical solution based on the finite-element method (1D FEM) computed with Openwind~\cite{Openwind2022}. All calculations refer to air at 20~$^\circ$C with physical properties given by Eq.(5.142) in Chaigne and Kergomard~\cite{Chaigne2016}.}}
\raggedright
\end{figure}

We demonstrate the differences between the cone models for the first impedance peak frequency $f_{R1}$. This is an essential property for sound generation in wind instruments, where the excitation mechanism is non-linearly coupled to the resonating air column. 
\\
A numerical comparison is carried out using the same computation scheme, by modifying the parameters in Eq.~(\ref{eq:TM_Nederveen}) as follows:
The model of Kulik~\cite{Kulik2007} is obtained by setting $\alpha_{fi} = \alpha_{gi}=f_i = g_i= 0$ and replacing $k\,x_i$ by $\omega/c\,(1+\alpha^\prime/r_i(1-\mathrm{j}))\,x_i$ with $i = 1,2$.
Another model is the \emph{cylindrical-slices model}. Additional to the above changes, it approximates the cone by a chain of $n$ discrete cylinders. The transfer matrix of this geometry is the product of $n$ transfer matrices according to Eq.~(\ref{eq:TM_Nederveen}) where $r_{1,j} = r_{2,j} = r_{\mathrm{eff},j}$ are replaced by the discrete radii $r_j$ ($j =1\ldots n$) approximating $r(x)$.
Such a model can be regarded as a reference for cones with weak taper $m$, as typically found in woodwind air columns. The area ratio between planar and spherical wavefronts $c_m$ is $1+\mathcal{O}(m^2)$, and so the bulging of the wavefront is negligible in the cases studied here.
For straight conical frustums representing simplified geometries of woodwinds (see Table~\ref{tab:woodwinds}), Fig.~\ref{fig:FIG1} shows the fundamental $f_{R1,con}$ of the two single-cone models relative to the fundamental $f_{R1,cyl}$ of the \emph{cylindrical-slices model}. For each of these straight cones, the length $l$ is varied by cropping from both ends while the taper remains constant. 
This shows that both single-cone models of Kulik and Nederveen converge towards the \emph{cylindrical-slices model} for small $l$, whereas the deviation becomes significant for long, slim cones.\\
The minor but systematic differences between the models raise the question of a reliable reference solution.
The theory of Zwikker and Kosten~\cite{Zwikker1949} captures the physical problem well but needs a discretization along the length. 
To compare the transmission matrix models to this reference, we use a very long slim conical frustum as a test case, with the taper and input radius of a bassoon, and 3~m length closed at its far end. \\
In Fig.~\ref{fig:FIG2}, the results of the two single-cone models by Kulik~\cite{Kulik2007} and Nederveen~\cite{Nederveen1969}  [Eq.~(\ref{eq:TM_Nederveen})], the \emph{cylindrical-slices model}, and the model by Zwikker and Kosten~\cite{Zwikker1949} are compared. The latter is computed using one-dimensional (1D) finite-elements with the software OpenWind~\cite{Openwind2022}. 

\begin{table}[htbp]
\centering
\begin{tabular}{lllll}
			& bassoon & oboe & tenor sax & soprano sax  \\ \hline
 $R_1$ [mm]& 2.0 & 1.5 & 4.1 &3.5\\
 $L$  [mm]& 2501 & 566  & 1369 & 650\\
 $m$  [mm/m]& 7 & 12.4 & 26.5 & 35.5
\end{tabular}
\caption{\label{tab:woodwinds}Conical frustum geometries ($R_1$, input radius; $L$, length; $m$ taper) related to typical woodwinds~\cite{Nederveen1969}}
\end{table}

It can be observed that Nederveen's model provides a better agreement to a discretized numerical reference solution compared to the other approaches mentioned above. The conceptual difference between the two single-cone models is that Nederveen~\cite{Nederveen1969} builds on a visco-thermal spherical wave equation, while Kulik~\cite{Kulik2007} builds on the solution for a loss-free cone and subsequently includes the loss model of a cylinder. 

\section{Conclusion}
A common approach to account for wall losses in a 1D model of a cone is to integrate the local propagation constant of a cylinder along the cone's center axis.
Compared to a sequence of cylindrical slices, this approach predicts slightly larger damping. 
This overestimation of the dissipation is not due to the bulging of the wavefront, as it also appears for weakly tapered cones with almost perfectly planar wavefronts.
The deviation is generally very small but increases with length, taper, and decreasing input radius and can become significant for long slim cones (Fig.~\ref{fig:FIG1}).
The derivation of the models has been reviewed by Thibault et al.~\cite{Thibault2023}, who point out that no exact closed-form solution for the governing equations exists.\\
Here, we rely on an approximate solution already derived by Nederveen~\cite{Nederveen1969}, which probably was the first attempt to include visco-thermal losses in an analytical description of the acoustic wave propagation in a cone.
This pioneering work agrees remarkably well with an up-to-date finite-element model for a long slim cone (Fig.~\ref{fig:FIG2}), without the need for discretization along the axis at the expense of evaluating two sine and cosine integral expressions~[Eq.~(\ref{eq:fg})]. In the form of a transfer matrix [Eq.~(\ref{eq:TM_Nederveen})], it can directly be included into existing impedance calculation frameworks. Our code is publicly available~\cite{ConeTransferMatrix2023}.\\
Future work could focus on convergence and consistency of this transfer matrix to ensure a smooth transition towards the cylindrical geometry and to study reciprocity.
 
 \section*{Acknowledgments}
The authors thank Alexis Thibault, Augustin Ernoult, Juliette Chabassier, Malte Kob, and Peter K\"oltzsch for valuable discussions and support. Augustin Ernoult kindly provided the results of the finite-element calculations.

\end{document}